\documentclass{iopart} 

\usepackage{enumerate}

\expandafter\let\csname equation*\endcsname\relax
\expandafter\let\csname endequation*\endcsname\relax
\usepackage{amsmath, amsthm, amssymb, amsfonts}

\usepackage{graphicx}
\usepackage[colorlinks=true,linkcolor=blue,citecolor=blue,filecolor=cyan,urlcolor=blue,breaklinks=false]{hyperref}

\usepackage{cite}

\newcommand{\mean}[1]{\left\langle #1 \right\rangle}
\renewcommand{\vec}[1]{{\boldsymbol{#1}}}
\renewcommand{\r}{\vec{r}}
\newcommand{\n}{\vec{n}}
\newcommand{\ei}{\vec{e}_i}
\newcommand{\Da}{D_{\rm {ac}}}
\newcommand{\Dt}{D_{\rm {tr}}}
\newcommand{\sa}{\sigma_{{\rm ac}}}
\newcommand{\sd}{\sigma_{{\rm tr}}}
\newcommand{\bzeta}{\vec{\zeta}}
\newcommand{\bnabla}{\vec{\nabla}}
\renewcommand{\u}{u_\mathrm{ac}}
\renewcommand{\th}{_{\rm{tr}}}
\newcommand{\ac}{_{\rm{ac}}}
\newcommand{\tot}{_{\rm{tot}}}

\begin{document}
\title{Entropy production of active particles and for particles in active baths}
\author{Patrick Pietzonka and Udo Seifert}

\address{II. Institut f\"ur Theoretische Physik, Universit\"at Stuttgart, 70550 Stuttgart, Germany}

\begin{abstract}
  Entropy production of an active particle in an external potential is
  identified through a thermodynamically consistent minimal lattice model that
  includes the chemical reaction providing the propulsion and ordinary
  translational noise. In the continuum limit, a unique expression follows,
  comprising a direct contribution from the active process and an indirect
  contribution from ordinary diffusive motion.  From the
  corresponding Langevin equation, this physical entropy production cannot be
  inferred through the conventional, yet here ambiguous, comparison of
  forward and time-reversed trajectories.  Generalizations to several
  interacting active particles and passive particles in a bath of active ones
  are presented explicitly, further ones are briefly indicated.
\end{abstract}
\vspace{1cm}

Active particles have become a major paradigm of non-equilibrium statistical
physics. Endowed with some propulsion mechanism, they perform persistent motion
in one direction until the latter changes either due to smooth rotational
diffusion or, as in bacteria, due to major reorientation. Already on the
single particle level and even more so for many interacting ones, they exhibit
a variety of new phenomena, which can be modeled and understood with effective
equations of motion as described in several recent reviews
\cite{rama10,cate12,roma12,marc13,elge15,cate15,bial15,pros15,bech16,star16,osha17,illi17,rama17}. A
related class of systems are those where passive particles are put into a bath
of such active ones, like colloidal particles in a laser trap that are pushed
around by bacteria present in the embedding aqueous solution
\cite{kris16,argu16}. On a phenomenological level, the motion of the passive
particle in such a non-equilibrium bath
\cite{chen07,vand15,magg14,stef16,wulf17,zaki17} can be described in a quite
similar way as in the first case.

Several recent works seek to formulate a thermodynamic foundation for the so
far phenomenological models of active Brownian particles
\cite{gang13,chau14,marc15,chak16,spec16a,spec17}.  In particular, identifying
the entropy production necessarily associated with any such non-equilibrium
system is a non-trivial problem. One route is to follow stochastic
thermodynamics \cite{seif12} and identify entropy production from comparing
the weights for forward and time-reversed trajectories. However, such a
procedure depends on the chosen level of coarse-graining and is hence not
unique, as recent studies of active Ornstein-Uhlenbeck particles have shown
\cite{fodo16,mand17,marc17,pugl17}. Similar concepts have been applied in
order to identify entropy production for field theories of active matter
\cite{nard17} and for active particles driven in linear response by chemical
reactions \cite{gasp17}.

In this paper, we identify entropy production for active particles by first
starting from a simplified but still thermodynamically consistent lattice model
where a chemical reaction leads to propulsion. Earlier lattice models for
active particles have not considered this thermodynamic aspect
\cite{thom11,solo13,slow16}. In addition to this active process, we allow for
re-orientations of the director and include ordinary translational diffusion.
In the continuum limit of a small step size per reaction event, we find a
unique expression for the total entropy production.  On the
basis of an effective Langevin description of the model, however, two different
definitions of entropy production based on the weights of forward and
time-reversed trajectories are conceivable. We demonstrate that these differ from the
total entropy production and trace this discrepancy to the fact that any such
Langevin equation contains implicit coarse-graining as soon as at least two
microscopic processes govern the motion like here chemical reactions and the
ever-present thermal translational noise.

\begin{figure}
  \centering
  \includegraphics[width=0.5\textwidth]{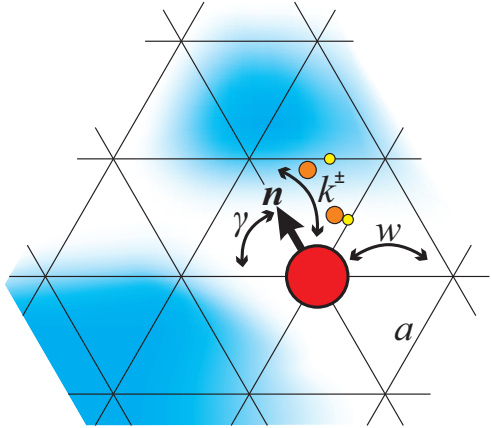}
  \caption{Illustration of the lattice model for an active particle. The
    particle jumps along the edges of a lattice in a potential landscape.
    Transitions along the director $\n$ can be driven by a chemical reaction
    (indicated by the rates $k^+$ and $k^-$).  Thermal noise induces rotational transitions
    of the director $\n$ (rate $\gamma$) and translational transitions in
    any lattice direction (rate $w$).}
  \label{fig:latticemodel}
\end{figure}
We start with a lattice model for an active particle in two dimensions on
either a square or a triangular lattice with lattice constant $a$, see
Fig.~\ref{fig:latticemodel}.  The sites are labeled by $\r_i$ and the four
(six) lattice vectors oriented along the adjacent edges of the square
(triangular) lattice are labeled by $\ei$.  The active particle has an orientation $\n$ that is at
any time parallel to one of these lattice vectors. Propulsion of the particle
one lattice unit in direction $\n$ arises from a not further specified
chemical reaction that liberates a free energy $\Delta \mu$ in each reaction
step.  This chemical reaction occurs with a rate $k^+$ in this forward
direction. Thermodynamic consistency requires that the rate $k^-$ for the
backward reaction with a concomitant step in the $-\n$ direction to happen
obeys the ratio
\begin{equation}
k^+/k^-=\exp\Delta \mu .
\end{equation} Throughout the paper we set $k_B=T=1$, i.e., entropy is dimensionless and
all energies are measured in units of the embedding bath temperature. If the
orientation $\n$ of the particle was fixed, the mean velocity $\u$ and the dispersion
$\Da$ of this then effectively one-dimensional motion would be given by
\begin{equation}
\u=(k^+-k^-)a ~~~\textup{and}~~~ \Da=(k^++k^-)a^2/2.
\end{equation}
It will later be  convenient to express the rates in terms of
these quantities as
\begin{equation}
k^\pm=\Da/a^2\pm \u/2a .
\end{equation}

In the presence of an additional potential $V(\r_i)$, 
 thermodynamic consistency requires that these rates 
become inhomogeneous. The symmetric choice
\begin{equation}
k^\pm(\r_i,\n)=k^\pm\exp[-(V(\r_i\pm a\n)-V(\r_i))/2]
\label{eq:rates}
\end{equation}
ensures that the local detailed balance condition 
\begin{equation}
k^+(\r_i,\n)/k^-(\r_i+a\n,\n)=\exp[V(\r_i)-V(\r_i+a\n)+\Delta\mu]
\end{equation}
is respected.

In addition to these moves generated by the chemical reaction driving the
particle, the orientation $\n$ becomes stochastic as well. In the simplest
case, this orientation undergoes Markovian transitions with a rate $\gamma$
between two adjacent orientations.  Under this dynamics, for periodic boundary
conditions, or with a confining potential on an infinite plane, the particle
reaches a steady state with the probability $p(\r_i,\n)$ to find it at site
$\r_i$ with an orientation $\n$. The stationary probability current from
$\r_i$ in the direction of $\n$ is then
\begin{equation}
  \label{eq:jacdisc}
  j\ac(\r_i,\n)\equiv p(\r_i,\n)k^+(\r_i,\n)-p(\r_i+a\n,\n)k^-(\r_i+a\n,\n).
\end{equation}

The entropy production rate $\sa$ of this active particle in the steady state
now follows uniquely from the usual rules of stochastic thermodynamics by
summing over all transitions as \cite{schn76,seif12}
\begin{equation}
\sa=\sum_{i,\n}{j\ac(\r_i,\n)}\ln[k^+(\r_i,\n)/k^-(\r_i+a\n,\n)] .
\label{eq:sa}
\end{equation}
There is no explicit contribution from the re-orientational moves of the
director $\n$, since the forward and backward rate for each such link
is the same, given by $\gamma$, and hence the corresponding logarithmic term would vanish. This
orientational dynamics, however, affects the stationary state,
 which is, in general,
hard to compute but not explicitly required for stating this expression of
entropy production due to the active process. 

Preparing for a continuum limit, we now assume that the lattice constant $a$ is
small compared to the length-scale of variations of the potential and the
stationary distribution. We can then expand the rates \eqref{eq:rates} as
\begin{equation}
k^\pm(\r_i,\n)\approx (\Da/a^2)(1\pm \u a/2\Da)(1\mp a\n\bnabla V(\r_i)/2) .
\end{equation}
Likewise, the stationary distribution can be expanded as
\begin{equation}
p(\r_i+a\n,\n)\approx p(\r_i,\n)[1+a\n \bnabla\ln p(\r_i,\n)].
\end{equation}
Inserting these expansions into \eqref{eq:jacdisc} and \eqref{eq:sa}
and collecting the terms that dominate in the limit $a\to0$, one gets
\begin{equation}
  \label{eq:ja}
  j\ac(\r_i,\n)=p(\r_i,\n)[\u-\Da\n\bnabla (V(\r_i)+\ln p(\r_i,\n))]/a
\end{equation}
for the magnitude of the corresponding vector-valued current density
$\vec{j}\ac(\r_i,\n)\equiv\n j\ac(\r_i,\n)$.  The entropy production rate becomes
\begin{equation}
\sa= \sum_{i,\n}{j\ac(\r_i,\n)}\,a\,[\u/\Da- \n\bnabla V(\r_i)] .
\end{equation} 
In the continuum limit, we implicitly redefine $p(\r,\n)$ and $\vec{j}(\r,\n)$
as densities by dividing $p(\r_i,\n)$ and $\vec{j}(\r_i,\n)$ by the area of
the associated unit cell of the lattice; $\vec{j}(\r_i,\n)$ obtains the
correct dimensionality of a current density by additional multiplication with
$a$.  Replacing now the summation over $\r_i$ by an integration over $\r$, we
find as our first result that the entropy production rate due to the active
process can finally be expressed as
\begin{align}
\sa&=\langle(\u- \Da \n\bnabla V(\r))^2\rangle/\Da -\Da\sum_\n\int d\r
[\n\bnabla p(\r,\n)][\u-\n\bnabla V(\r)]\nonumber\\
&=\langle(\u- \Da \n\bnabla V(\r))^2\rangle/\Da -\Da \langle(\n\bnabla)^2
V(\r)\rangle,
\label{eq:sa2}
\end{align}
where we have applied integration by parts to the second term.  Here, and in
the following, the angular brackets $\langle ... \rangle$ denote an average in
the stationary state $p(\r,\n)$.

So far, we have not yet included ordinary thermal translational diffusion. We can
do so by allowing additional moves from $\r_i$ to $\r_i+a\ei$ with a rate 
\begin{equation}
w(\r_i,\ei)=w \exp[-(V(\r_i + a\ei)-V(\r_i))/2],
\end{equation}
 where $w\equiv \Dt a^2
$ is the bare discrete jump rate and $\Dt$ the  diffusion
constant on a square lattice. For a triangular lattice, one has to choose
$w=2\Dt/3a^2$. 
The corresponding probability current from $\r_i$ in the direction $\ei$ is then
\begin{equation}
  \label{eq:jtrdisc}
  j\th(\r_i,\n,\ei)=p(\r_i,\n)w(\r_i,\ei)-p(\r_i+a\ei,\n) w(\r_i,-\ei).
\end{equation}
Going through the same steps as
above, for both lattices, one gets the current density
\begin{equation}
  \label{eq:jtr}
  \vec{j}\th(\r_i,\n)=-\Dt p(\r_i,\n)\bnabla [V(\r_i)+\ln p(\r_i,\n)]
\end{equation}
and the entropy production rate due to the translational diffusive steps
\begin{equation}
\sd = -\sum_\n \int d\r \,\vec{j}\th(\r,\n)\bnabla V(\r) = \Dt\langle [\bnabla V(\r)]^2 -\bnabla^2 V(\r)\rangle ,
\label{eq:sd}
\end{equation} which is, in the first form, well-known from ordinary
stochastic thermodynamics as the entropy production in the medium
\cite[Sec. 2.5]{seif12}, which is in the steady state equal to the total
entropy production, and the second 
equality follows with an integration by parts.
In equilibrium, i.e., without an active process, $\sd$ vanishes since for a
Boltzmann distribution $\sim \exp[-V(\r_i)]$ the two terms above cancel. For an active process,
the stationary distribution is no longer of this form and, hence, in the
presence of a potential even these translational 
diffusive steps can contribute to the entropy production rate, which is our
second main insight. The total entropy production rate becomes
\begin{align} 
\sigma\tot&\equiv \sa+\sd\nonumber\\
 &=\langle(\u- \Da \n\bnabla V(\r))^2\rangle/\Da -\Da \langle(\n\bnabla)^2
V(\r)\rangle+\Dt\langle [\bnabla V(\r)]^2 -\bnabla^2 V(\r)\rangle.
\label{eq:stot}
\end{align} 

Since this result no longer contains any explicit reference to the originally
underlying lattice, it should be obvious that it holds as a continuum result
also for three-dimensional active motion. Moreover, it holds  for any 
non-driven Markovian dynamics of the orientation $\n$
as long as the latter is independent of the spatial position. Therefore, it
includes ordinary  active Brownian particles with their orientational
diffusion \cite{roma12}.

The stationary distribution underlying above averages follows in the discrete
case from the stationary master equation, which we cast in the shape of a
balance equation for the edge currents \eqref{eq:jacdisc} and \eqref{eq:jtrdisc}
\begin{equation}
  0=\partial_t
  p(\r_i,\n)=j\ac(\r_i-a\n,\n)-j\ac(\r_i,\n)-\sum_{\ei}j\th(\r_i,\n,\ei)+L_\n p(\r_i,\n),
\end{equation}
where the operator $L_{\n}$ acting solely on $\n$ conveys the non-driven
dynamics of the rotational degree of freedom. Likewise in the continuum limit,
the stationary distribution is the solution of the stationary Fokker-Planck
equation that follows from the continuity of the currents \eqref{eq:ja} and \eqref{eq:jtr} as
\begin{align}
\label{eq:fokkerplanck}
  0=\partial_t p(\r,\n)&=-\bnabla[\vec{j}\ac(\r,\n)+\vec{j}\th(\r,\n)]+L_{\n} p(\r,\n)\nonumber\\
&=-\bnabla\n[\u-\Da\n(\bnabla
  V(\r))]p(\r,\n)+\Da(\n\bnabla)^2p(\r,\n)
  \nonumber\\
  & \qquad +\Dt\bnabla[(\bnabla
  V(\r))p(\r,\n)]+\Dt\bnabla^2p(\r,\n)+L_{\n} p(\r,\n).
\end{align}
For a continuous space of orientations $\n$, the operator $L_\n$ becomes the
Fokker-Planck operator governing the stochastic dynamics of $\n$. For example,
in two dimensions, for a dipolar interaction through the potential $\mathcal{U}(\n)=-\vec{B}\n$ , this operator would read
$L_\n\propto \partial_\theta^2+B\partial_\theta\sin\theta$, where
$\theta$ denotes the angle between $\n$ and the homogeneous field~$\vec{B}$.
Multiplication of equation \eqref{eq:fokkerplanck}
by $V(\r)$ and repeated integration by parts leads to the relation
\begin{equation}
  0=\mean{\n\bnabla V(\r)[\u-\Da\n\bnabla V(\r)]}+\Da\mean{(\n\bnabla)^2V(\r)}-\Dt\mean{[\bnabla
    V(\r)]^2}+\Dt\mean{\bnabla^2V(\r)},
\end{equation}
which can be used to write the expression \eqref{eq:sa2} for the active
part of the entropy production as
\begin{equation}
\sa=\u^2/\Da -\u\mean{\n\bnabla V(\r)}-\Dt\mean{[\bnabla V(\r)]^2}+\Dt\mean{\bnabla^2 V(\r)}.
\label{eq:sa3}
\end{equation}
Consequently, the total entropy production \eqref{eq:stot} becomes
\begin{align} 
\sigma\tot=\u^2/\Da -\u\mean{\n\bnabla V(\r)}.
\label{eq:stot2}
\end{align} 

For a comparison with previous work, we state for our model on the continuum level
the corresponding 
Langevin dynamics, which can be read off from the Fokker-Planck equation \eqref{eq:fokkerplanck} as
\begin{equation}
\dot{\r}=\u\n -(\Dt+\Da\n\otimes\n)\bnabla V(\r) + \bzeta\th + \zeta\ac\n .
\label{eq:langevin}
\end{equation} 
It contains ordinary thermal translational noise with the correlations
\begin{equation}
\langle \bzeta\th(t_2)\otimes\bzeta\th(t_1)\rangle = 2\Dt\boldsymbol {1}
\delta(t_2-t_1).
\label{eq:wnth}
\end{equation} 
 The chemical driving leads to an active noise, which acts along
$\n$, has correlations
\begin{equation}
\langle  \zeta\ac(t_2)\zeta\ac(t_1)\rangle = 2 \Da \delta(t_2-t_1),
\label{eq:wnac}
\end{equation}
 and is uncorrelated with the
translational noise. Thermodynamic consistency thus leads to an active
``mobility'' $\Da \n\otimes\n$ for motion in a potential, which is reminiscent
of the coupling described in Ref.~\cite{gasp17} in linear response.

There is no need to specify the orientational dynamics for $\n(t)$ explicitly, provided it is Markovian, independent of $\r$, and non-driven. Under these
conditions, the stationary distribution factorizes $p(\r,\n)=p(\r|\n)p(\n)$
with
\begin{equation}
p(\n)\sim \exp[-\mathcal{U}(\n)],
\end{equation}
where the potential $\mathcal{U}(\n)$ is constant for ordinary rotational diffusion and
$\mathcal{U}(\n)=-\vec{B}\n$ if the director is subject to a dipolar interaction
with a field $\vec{B}$.

It might look tempting to infer the entropy production directly from
the Langevin equation \eqref{eq:langevin} by comparing the weight of a forward trajectory
$\r(t),\n(t)$ (for $0\leq t\leq \mathcal{T}$)
with that of the corresponding backward trajectory $\tilde \r(t)=\r(\mathcal{T}-t),
\tilde \n(t)=\n(\mathcal{T}-t)$ \cite{maes03b,seif12}. This apparently innocent procedure fails in the
presence of two noise sources as the following simple example of free
active motion shows. The Langevin equation
\begin{equation}
\dot{\r}=\u\n + \bzeta\th + \zeta\ac\n
\end{equation} 
with the above given noise correlations leads to the 
weight (conditioned on the $\n$-trajectory) 
\begin{equation}
p[\r(t)|\n(t)]\sim\exp\int_0^\mathcal{T} dt\bigg\{-\frac{(\dot{\r}\n-\u)^2}{4D_{||}}-\frac{[(\boldsymbol{1}-\n\otimes\n)\dot{\r}]^2}{4\Dt}\bigg\},
\label{eq:pathweight}
\end{equation} 
which follows from the path weights of the components of the white noise
(\ref{eq:wnth}, \ref{eq:wnac}) in the directions parallel and perpendicular to
$\n$. For the former, the diffusion coefficients for the active and the
translational noise add up to $D_{||}\equiv \Dt+\Da$.  The part of the
exponent that is asymmetric under time reversal becomes after averaging
\begin{equation}
\varSigma \equiv 
\mean{\ln\frac{p[\r(t)|\n(t)]}{p[\tilde{\r}(t)|\tilde{\n}(t)]}}/\mathcal{T} = 
\u^2/(\Dt+\Da) \leq \sigma\tot=\u^2/\Da,
\label{eq:pathratio}
\end{equation}
which is less than what we have identified above as physical entropy production rate $\sigma\tot$. 
This conventional procedure fails in this case since the
trajectory no longer contains the full thermodynamically relevant information.
An increment along the $\n$-direction could be either due to thermal translational
noise (without concomitant entropy production in the absence of a potential)
or due to an active chemical reaction which comes with entropy production.
A comparison on the level of forward and backward trajectories thus implicitly
involves some coarse-graining of the underlying microscopic processes in which case the
true entropy production is underestimated by a \textit{bona-fide} comparison of the
weights \cite{kawa07,gome08b,rold10,mehl12,bo14,bo17}.

The inequality $0\leq\varSigma\leq\sigma\tot$ holds even in the presence of
the potential $V(\r)$, as we now show. The conditioned path weights
corresponding to the Langevin equation~\eqref{eq:langevin} read
  \begin{align}
    p[\r(t)|\n(t)]\sim\exp\int_0^\mathcal{T} dt\bigg\{&-\frac{[\dot{\r}\n-\u+D_{||}\n\bnabla
      V(\r)]^2}{4D_{||}}-\frac{[(\boldsymbol{1}-\n\otimes\n)(\dot{\r}+\Dt\bnabla
      V(\r))]^2}{4\Dt}\nonumber\\
&+\frac{D_{||}}{2}(\n\bnabla)^2V(\r)+\frac{\Dt}{2}[(\boldsymbol{1}-\n\otimes\n)\bnabla]^2V(\r)
\bigg\},
  \end{align}
  where the terms in the second line arise from the
  Stratonovich discretization rule. With this expression, the quantity
  $\varSigma$ follows as
  \begin{align}
    \varSigma &\equiv \mean{ \ln\frac
{p[\r(t)|\n(t)]}{p[\tilde{\r}(t)|\tilde\n(t)]}}/\mathcal{T}\nonumber\\
&=\mean{(\dot{\r}\n)\u/D_{||}-\dot{\r}\bnabla
  V(\r)}
=\u\mean{\dot{\r}\n}/D_{||}=\u^2/D_{||}-\u\mean{\n\bnabla V(\r)}.
\label{eq:Sigma}
  \end{align}
Here, we use that the average $\mean{\dot{\r}\bnabla V(\r)}$ vanishes, since
it is the total
differential of $V(\r)$, and evaluate the average $\mean{\dot{\r}\n}$ via integration by parts from $\int d\r d\n \vec{j}(\r,\n)\n$ with the
steady state current
\begin{equation}
  \vec{j}(\r,\n)\equiv [\Da\n(\u/\Da-\n(\bnabla V(\r)))-\Da\n(\n\bnabla)-\Dt(\bnabla V(\r))-\Dt\bnabla]p(\vec r,\vec n).
\end{equation}
In the fictitious, dynamically equivalent, process for which active and translational
diffusion in $\n$-direction are attributed to the same source of noise,
$\varSigma$ would correspond to the true thermodynamic entropy production and
therefore $\varSigma\geq 0$. On the other hand, the positivity of $\Dt$
directly shows that  $\varSigma\leq \sigma\tot$ in the comparison of
Eq.~\eqref{eq:Sigma} with \eqref{eq:stot2}.
Equality, $\varSigma=\sigma\tot$, is reached only in the limit of vanishing translational
noise \cite{pugl17}, i.e., for $\Dt\to 0$. 

For models of active particles that lack both ordinary and active translational diffusion, the full trajectory of the system is
described by $\r(t)$ whereas the director $\n$ becomes a dependent
variable. The time reversal used in Ref.~\cite{fodo16} to define an entropy
production rate therefore involves, in the simple case of a free particle, the
implicit reversal of the director $\n$. Inspired by this definition, we now consider
for the present model a ratio of path weights where this reversal is made
explicit, leading to the quantity
  \begin{equation}
    \label{eq:Sigmaprimedef}
    \varSigma'\equiv \mean{ \ln\frac
      {p[\r(t)|\n(t)]}{p[\tilde{\r}(t)|-\tilde\n(t)]}}/\mathcal{T}=\u\mean{\n\bnabla V(\r)},
  \end{equation}
which is zero for a vanishing potential, as is also the case for the entropy production
defined in Ref.~\cite{fodo16}. Interestingly, this quantity adds up
with $\varSigma$ to the
potential-independent constant
\begin{equation}
  \varSigma+\varSigma'=\u^2/D_{||}
  \label{eq:sumrule}
\end{equation}
and likewise with the total entropy production to $\sigma\tot+\varSigma'=\u^2/\Da$ .
If the dynamics of the director $\n$ is symmetric under flipping
$\n\mapsto -\n$, i.e., $p[-\n(t)]=p[\n(t)]$,  and non-driven, the quantity $\varSigma'$
satisfies $0\leq\varSigma'\leq \u^2/D_{||}$, due to the positivity of
$\varSigma$ and Jensen's inequality in
\begin{equation}
  \varSigma'=-\mean{ \ln\frac{p[\tilde{\r}(t),-\tilde\n(t)]}{p[\r(t),\n(t)]}}/\mathcal{T}\geq
   -\ln\mean{\frac{p[\tilde{\r}(t),-\tilde\n(t)]}{p[\r(t),\n(t)]}}/\mathcal{T}
   =0.
\end{equation}
In spite of this positivity, $\varSigma'$ does not qualify as a
characterization of the physical entropy production, since its dependence on a
potential (for fixed $\u$ and $\Da$) is complementary to the physical
dissipation (i.e., to $\sigma\tot$).
The distinction between $\varSigma$ and $\varSigma'$  is reminiscent of a
similar alternative in the case of external flow \cite{spec08}, see also \cite{spec17}.

The ``sum-rule'' \eqref{eq:sumrule}, however, applies only on the continuum
level. In the discrete model, the contribution to the logarithmic ratio of
path weights \eqref{eq:Sigmaprimedef} from transitions amounts to only the
respective changes $\Delta V$ of the potential, since the contribution
$\Delta\mu$ from the chemical reaction cancels for inverted $\n$ in the
backward path. Since the change of the potential is zero on average, the only
contribution to $\varSigma'$ stems from the sojourn times in states where the
exit rates for forward and backward path do not cancel, leading to
  \begin{equation}
    \varSigma'=\sum_{i,\n}p(\r_i,\n)[-k^+(\r_i,\n)-k^-(\r_i,\n)+k^+(\r_i,-\n)+k^-(\r_i,-\n)],
    \label{eq:sc}
  \end{equation}
with no obvious relation to $\varSigma$ or $\sigma\tot$.

So far, we have dealt with one active particle. Our approach is easily generalized to $N$ interacting particles
with positions $\r^j$ and orientation $\n^j$ leading to
\begin{equation}
\sa =\sum_{j=1}^N\left\{\langle(\u- \Da\n^j \bnabla^j V(\{\r^j\}))^2\rangle/\Da -\Da \langle(\n^j\bnabla^j)^2 V(\{\r^j\})\rangle\right\} ,
\end{equation} where $V(\{\r^j\})$ comprises the two-body interaction potential 
and a possible external
one-body potential and $\bnabla^j\equiv \bnabla_{\r^j}$. The diffusive
translational contribution $\sd$ \eqref{eq:sd} is likewise additive in the $N$ particles.

This formalism can also easily be applied to $M$ passive colloidal particles
($j=N+1,\dots,N+M$) in  
a non-equilibrium bath of $N$ active ones ($j=1,...,N$) leading to
\begin{align}
\sigma\tot
&=\sum_{j=1}^N\left\{\langle(\u- \Da\n^j \bnabla^j V(\{\r^j\}))^2\rangle/\Da -\Da \langle(\n^j\bnabla^j)^2 V(\{\r^j\})\rangle\right\}\nonumber\\
&\qquad+ \sum_{j=1}^{M+N}\left\{\langle \Dt^j \bnabla^j V(\{\r^j\})^2\rangle
  -\Dt^j \langle(\bnabla^j)^2 V(\{\r^j\})\rangle\right\}\nonumber\\
&=N\u^2/\Da-\u\sum_{j=1}^N\mean{\n^j\bnabla^jV(\{\r^j\})},
\label{eq:bathent}
\end{align}  
where $V(\{\r_i\})$ includes all two-body interactions and the external potential.

\begin{figure}
  \centering
  \includegraphics[width=0.7\textwidth]{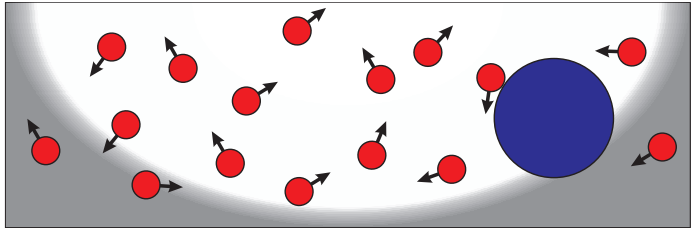}
  \caption{Illustration of a passive particle (blue) interacting via hard-core
    repulsion with many active particles (red). An external harmonic potential
  (indicated in gray) acts only one the passive particle.}
  \label{fig:activetrap}
\end{figure}
For a simple example, assume that an external harmonic potential of strength $k$
acts only on one passive colloid embedded in a bath of active ones, see Fig.~\ref{fig:activetrap}, as
realized in experiments \cite{kris16,argu16}. All particles (passive and active ones) 
are hard-core repulsive. From the first equality in \eqref{eq:bathent}, the total entropy
production then becomes 
\begin{equation}
\sigma\tot= \rho N\u^2/\Da +  \Dt k(k\langle \r^2\rangle -d) 
\end{equation}
 in $d$ dimensions and with $\r$ denoting the position of the passive colloid. On the
lattice level, one has to correct for the fact that
 the reaction step is possible only if the
neighboring lattice site is unoccupied which happens for a fraction $\rho<1$ 
of total phase space that is, in general, not easily computable. The second term is essentially
the net heat dissipated through the viscous friction of the passive
particle while  sliding down diffusively the potential after having been
pushed upwards by the active particles. 

Building on the minimal thermodynamically consistent model for active
particles presented above, several generalizations towards more detailed
models are straightforwardly conceivable. First, for non-spherical particles, the
translational diffusion tensor will typically not be isotropic. For rod-like
particles oriented along $\n$, one would replace $\Dt$ by a diffusion tensor of
the form
  \begin{equation}
    \mathbf{D}_\mathrm{tr}=\Dt^{||}\n\otimes\n+\Dt^\perp(\boldsymbol{1}-\n\otimes\n)
  \end{equation}
with a perpendicular component $\Dt^\perp$ and a parallel component $\Dt^{||}$.
The latter is then relevant in expressions containing $\Da+\Dt$. Second, one could allow for different reaction channels indexed by $\rho$,
each with a different $\u^\rho$ and $\Da^\rho$ and then sum over the $\rho$.

Third, one can add further internal degrees of freedom to the model, which act
as a memory and affect the preferred chemical reaction channel \cite{piet16a}. Formally, if
these internal degrees of freedom satisfy a non-driven Markovian dynamics,
they can be treated analogously to the degree of freedom associated with
$\n$. In particular, this generalization allows one to emulate the dynamics of
active Ornstein-Uhlenbeck particles using one continuous internal degree of freedom
$f$, which couples to the propulsion speed as $\u\propto f$ and whose
dynamics is such that $f\n$ describes an Ornstein-Uhlenbeck process.  

Fourth, one could replace the simple Poisson process used above for modeling the chemical
reaction by a more elaborate chemical reaction network, in which each
transition comes either with a reaction step or with a displacement step. 
Such a network would typically be multicyclic, allowing for a weak coupling
between the chemical reaction and the stepping of the particle. In a linear
response description of the network currents \cite{andr07b}, this coupling would
be expressed by a small Onsager coefficient \cite{gasp17}.

Popular models of active Brownian particles, for example the active
Ornstein-Uhlenbeck particles considered in Refs.~\cite{fodo16,mand17}, neglect
both active and thermal translational noise in the Langevin equation
\eqref{eq:langevin}. In our model, setting $\bzeta\th=0$ is thermodynamically
consistent. In that case, the entropy production $\varSigma$ in the effective
model becomes equal to the total entropy production $\sigma\tot$. In contrast,
setting $\zeta\ac=0$ breaks micro-reversibility, leading to trajectories that
lack a time-reversed counterpart. Thus, in the limit $\Da\to0$ with fixed
$\u$, the total entropy production $\sigma\tot$ diverges.  Likewise, in the
presence of a potential scaling like $V\sim 1/\Da$ and for $\Dt=0$, both
$\varSigma$ and $\varSigma'$ diverge.  Nonetheless, the Langevin equation
\eqref{eq:langevin} with $\bzeta\th=\zeta\ac=0$ may be a valid effective
description of large active particles with strong propulsion force. The
well-defined stationary distribution for this process can then be used for the
averages in Eq.~\eqref{eq:stot} or \eqref{eq:bathent} to obtain the leading
order contribution to $\sigma\tot$ for small but non-zero $\Da$. The
non-divergent entropies considered for active Ornstein-Uhlenbeck particles in
Refs.~\cite{fodo16,mand17} differ from $\sigma\tot$, $\varSigma$, and
$\varSigma'$ and rely on the variability of the propulsion speed $\u$ in order
to make the coarse-grained trajectories $\r(t)$ reversible. In models with
constant $\u$, as considered here, such a definition would not be
feasible, let alone lead to the thermodynamic entropy production
characterizing the dissipation in the surrounding medium.

In summary, we have determined entropy production for a large class of active
particles starting with a thermodynamically consistent lattice model. Its
continuum limit offers an alternative to attempts of defining entropy
production from path weights from an effective Langevin equation. These definitions
underestimate the total entropy production (whenever translational noise
cannot be ignored) and suffer from a certain arbitrariness when defining the
``time-reversed'' path.
We have focused on the stationary state and the average entropy production.
Extensions to relaxation from an arbitrary initial state or to time-dependent
potentials should be straightforward. Likewise, starting with the lattice model,
trajectory-dependent entropy production and the corresponding integral and 
detailed fluctuation theorems can be derived as in ordinary stochastic
thermodynamics \cite{seif12}.
Finally, it will be interesting, and presumably somewhat more challenging, to
extend this approach to hydrodynamic and field-theoretic descriptions of active
particles, and, more generally, of active matter \cite{bert09,nard17,illi17,sand17}.

\section*{References}

\bibliographystyle{bibgen}
\bibliography{/home/public/papers-softbio/bibtex/refs}

\end{document}